\documentclass[useAMS,usenatbib]{mnras}
\usepackage{ulem}
\usepackage{soul}
\usepackage{xcolor}
\usepackage{xspace} 

\usepackage{newtxtext}

\usepackage[T1]{fontenc}

\RequirePackage{fix-cm}

\usepackage{graphicx}	
\usepackage{amsmath}	
\usepackage{newpxmath}

\newif\ifrevision
\revisionfalse

\newcommand{\rev}[1]{\ifrevision\textcolor{cyan}{#1}\else#1\fi}
\newcommand{\del}[1]{\ifrevision\textcolor{red}{\sout{#1}}\else\fi}

\usepackage{orcidlink}

\newif\ifsecondrev 
\secondrevfalse

\newcommand{\revs}[1]{\ifsecondrev\textcolor{blue!60!green}{#1}\else#1\fi}
\newcommand{\dels}[1]{\ifsecondrev\textcolor{red}{\sout{#1}}\else\fi}

\title[Short title, max. 45 characters]{A jet formation model for astrophysical objects}

\author[Chun Xu]{Chun Xu$^{1}$\thanks{E-mail: chun.xuu@shao.ac.cn}\orcidlink{0009-0009-2507-5977}
\\
$^{1}$Shanghai Astronomical Observatory, Chinese Academy of Sciences, Shanghai 200030, China}

\date{Accepted XXX. Received YYY; in original form ZZZ}

\pubyear{\the\year{}}

\begin{document}
\label{firstpage}
\pagerange{\pageref{firstpage}--\pageref{lastpage}}
\maketitle

\begin{abstract}
We propose a unified model for jet formation applicable to active galactic nuclei, young stellar objects, and X-ray binaries. In this model, the binding energy released from the accretion disk is primarily stored as turbulence rather than being radiated away, leading to the formation of advection-dominated accretion flows. Near the central object, a thick accretion disk with funnel-like structures develops. Within the turbulent flows, the smallest stable blobs can be accelerated beyond the escape velocity through \rev{the combination of two mechanisms --- the Gaussian-like velocity distribution within the turbulence and} a mechanism involving the combined effects of inward pressure force and angular momentum conservation. These rapidly moving blobs may exit through the funnels, collectively forming two opposing jets. This model predicts that jets originate from the innermost region of the thick disk surrounding the central object. \rev{The formation of jet is directly related a parameter $\eta$ that describes the energy fraction stored in turbulence in units of the binding energy
of local Keplerian energy. $\eta > 0.5$ is a minimal condition for jet to form.} \rev{This model} \del{It} can be extended to account for jet formation in active galactic nuclei, young stellar objects, X-ray binaries, and other analogous astronomical systems.
\end{abstract}




\section{Introduction}

It is well known that many classes of astrophysical objects, including active galactic nuclei (AGNs), young stellar objects (YSOs), and X-ray binaries (XRBs), contain highly collimated jets, and the jet velocities are always of the order of the escape velocities from the central objects. The similarity of the jets found in very different astrophysical environments suggests that the formation mechanism of the jets must be the same \citep{Livio1999}. Lots of effort has been made by different researchers to explore the formation machinery of the jets, and some mechanisms have become very popular. These include: the Blandford-Znajek mechanism, which assumes that the spin energy of the black hole is extracted to drive the jets \citep{Blandford1977}; the Blandford-Payne mechanism, which assumes that the accretion disk makes its poloidal magnetic field rotate and drives the material out along the open field lines to form the jet \citep{Blandford1982}; and the YSO X-wind model, which is a magnetocentrifugal mechanism for launching jets and winds from young stellar objects \citep{Shu1994,Pudritz1983,Pudritz1986}. The Blandford-Znajek mechanism only works for spinning black holes but not for YSOs, so it is not universal. The Blandford-Payne mechanism and the YSO X-wind model share many similarities and are becoming a very popular base model in MHD simulations. However, both models face many problems, such as the origin of the magnetic field, jet power and mass flux uncertain, model stability, etc. \citep{Anderson2005, Lubow1994}.

Lynden-Bell \citeyearpar{Lynden-Bell1978} proposed a jet formation model based on a thick accretion disk, where two funnels can form above and below the central black hole on both sides of the disk, through which materials can be accelerated to form two jets. Although the energy source driving jet acceleration remains poorly constrained and is assumed to be linked to viscosity, the funnel concept provides a compelling framework for understanding jet collimation within the thick disk. Abramowicz and Piran \citeyearpar{Abramowicz1980} later developed a jet acceleration model involving radiation pressure within the thick accretion disk. Abramowicz and collaborators also proposed a slim disk model to address the thermal instability of radiation-dominated thin disks \citep{Abramowicz1988}. The slim disk serves as a bridge between the standard thin disk model \citep{Shakura1973} and later-developed advection-dominated accretion flows.
In 1984, however, Papaloizou and Pringle \citeyearpar{Papaloizou1984} demonstrated that thick accretion disks are unstable to low-order non-axisymmetric modes. Such instability, now known as the Papaloizou–Pringle instability (PPI), poses substantial challenges to the viability of stationary thick disks. 

Narayan and Yi \citeyearpar{Narayan1994} proposed the Advection-Dominated Accretion Flow (ADAF) model for low-luminosity accretion onto black holes. At low mass accretion rates, the gas becomes optically thin and two-temperature. Most viscous heat is advected inward with the flow rather than being radiated, making the accretion extremely radiatively inefficient. The flow is geometrically thick, hot, and subsonic, resembling a hot ion torus rather than a thin disk. This model successfully explained the low luminosities of Sgr A* and quiescent black hole binaries, where standard thin disks would vastly overpredict the observed emission. Blandford and Begelman \citeyearpar{Blandford1999} proposed the Advection-Dominated Inflow-Outflow Solution (ADIOS) as a modification to the standard ADAF picture. They argued that in low-luminosity accretion, the majority of the accreting mass is lost to winds rather than reaching the black hole due to the conservation laws of energy and angular momentum. Both ADAF and ADIOS models yield global solutions that permit outflows from the central object.

The fundamental difference between the thin disk and ADAF lies here. The thin disk radiates away its liberated binding energy, causing its total energy to fall below the escape threshold; as a result, no outflow can emerge from the vicinity of the central object. In contrast, due to its low radiation efficiency, ADAF retains the released energy as heat and advects it inward. Since the total energy is not significantly reduced, outflows may form and escape from the central object. This is analogous to a comet approaching the Sun and being able to return to infinity, as no energy is lost. 

In this paper we propose a jet formation model as follows. The released binding energy is primarily stored in turbulence rather than converted into heat. As a result, the temperature of the accretion flow is not too high, and radiation is very limited. Thus an ADAF is formed while the total pressure is the combination of thermal pressure and turbulent pressure. The difference in this model is that we do not need to assume a two-temperature plasma to keep the radiation efficiency low. In an ADAF-like model, a thick disk is expected to form with two funnels around the central object \citep{Narayan1995, Blandford1999}. Near the inner edge of the thick disk, the effective force from the pressure gradient points toward the central object \citep{Lynden-Bell1978, Frank2002}. An inward-rotating gas blob will gain energy due to the work done by pressure while keeping its angular momentum conserved. Some of the gas blobs may acquire total energy surpassing the escape energy, allowing them to exit through the funnels and form a twin jet.

The structure of this paper is as follows. In Section 2, we discuss the ADAF model with the inclusion of turbulence. Section 3 addresses the thick disk and funnel formation, largely following earlier work, followed by an analysis of the turbulent blob acceleration mechanism and jet formation \rev{and with a short summary of the model}. Section 4 examines jets in AGNs, YSOs, and XRBs, comparing observational features with our model. \revs{Section 5 presents some cautionary notes of the model.} Finally, Section 6 summarizes our results and conclusions.

\section{Accretion flows}
To study accretion onto a point-mass central object, one would ideally solve the full Navier-Stokes equations—an approach that remains analytically intractable. As a simplification, we adopt a steady-state, axisymmetric configuration in cylindrical coordinates, the equation reduces to \citep{Frank2002}:

\begin{equation}
-\frac{1}{\rho}\boldsymbol{\nabla} P - \boldsymbol{\nabla}\psi + \Omega^2 \mathbf{R} + \frac{\mu}{\rho}\nabla^2 \mathbf{u} = \mathbf{0}
\label{eq:momentum}
\end{equation}

Where $P$ is the pressure, $\rho$ is the density, $\Omega$ is the (Keplerian) angular velocity, $\mu$ is the dynamic viscosity, $\psi = -GM/(R^2+z^2)^{1/2}$ is the gravitational field of central object, while $\mathbf{R}$ is radius from axis and $\mathbf{u}$ is the velocity. The parameters are functions of $R$ and $z$.
From the radial and azimuthal components of equation (1), along with the continuity equation and energy equation, Narayan and Yi \citeyearpar{Narayan1994} deduced a self-similar solution (see also: \citealt{Blandford1999,  Abramowicz1995}). They assumed 2 terms in the energy equation, $Q^+$ stands for energy input per unit area due to viscosity, and $Q^-$ stands for the energy loss due to radiation. When $Q^+-Q^-$ is close to 0 then a thin disk solution \citep{Shakura1973} is obtained. When $Q^-$ is much smaller or close to 0 then ADAF is formed.
In advection-dominant case, if the normalized Bernoulli constant $b > 0$ then the accretion flows are unbound so can go to infinity \citep{Narayan1994}. They suggest that jets and outflows are related to the advection-dominant flows.

The formation of an advection-dominated flow requires either a very low radiative efficiency, as seen in two-temperature plasmas, or an extremely high, super-Eddington accretion rate at which radiation is unable to effectively carry away the energy \citep{Narayan1994, Narayan1995}. This dual requirement for the accretion flow to facilitate the development of an advection-dominated regime appears rather implausible. Given that neither a two-temperature plasma nor super-Eddington accretion is likely to occur in YSOs, the ADAF model cannot be meaningfully applied to them—an unfortunate constraint.

Here we propose a mechanism that may consistently drive the formation of an advection-dominated accretion flow and the outflows. We suggest that the pressure in equation (1) consists of two components: one is the thermal pressure as usual, and the other is the pressure due to turbulence, so it is replaced by:

\begin{equation}
P_{\text{tot}} = P + P_t 
\end{equation}
where $P = (\rho/\mu m_p) k_BT$ is the thermal pressure and $P_t = \rho v_t^2$ is the turbulence pressure, with $v_t$ being the rms turbulent velocity. In full turbulence, $v_t$ can reach local Keplerian speed $v_K$. The energy density per unit mass in turbulence is
$E_t = \int_{1/l}^{\infty} E(k) dk  = \frac{1}{2} v_t^2$,
where $E(k) = C_K \epsilon^{2/3} k^{-5/3}$ is the Kolmogorov spectrum with Kolmogorov constant $C_K \approx 1.5$, $\epsilon$ is the energy dissipation rate per unit mass, $k$ is the wavenumber, and $l$ is the size of the turbulent region \citep{Pope2000}.
Then the four equations (1)--(4) in Narayan and Yi \citeyearpar{Narayan1994} become:

\begin{equation}
\frac{d}{dR} \left( \rho R H v_R \right) = 0
\end{equation}

\begin{equation}
v_R \frac{dv_R}{dR} - \Omega^2 R = -\Omega_K^2 R - \frac{1}{\rho} \frac{d}{dR} \left[ \rho \left( c_s^2 + v_t^2 \right) \right]
\end{equation}

\begin{equation}
v_R \frac{d(\Omega R^2)}{dR} = \frac{1}{\rho R H} \frac{d}{dR} \left( \frac{\alpha \rho (c_s^2 + v_t^2) R^3 H}{\Omega_K} \frac{d\Omega}{dR} \right)
\end{equation}

\begin{equation}
\Sigma v_R \left( T \frac{ds}{dR} + \frac{1}{2} \frac{dv_t^2}{dR} \right) = Q^+ - Q^- 
\end{equation}

Here $v_R$ denotes the radial velocity, $c_s^2 = P/\rho$ represents the sound speed, $\Sigma = 2\rho H$ is the surface density, $H \sim R \sqrt{c_s^2 + v_t^2}/v_K$ gives the vertical scale height. The symbols $\Omega$ and $\Omega_K$ correspond to the angular velocity and Keplerian angular velocity, respectively, while $\alpha$ stands for the Shakura-Sunyaev viscosity parameter. The local temperature is denoted by $T$, and $s$ in (6) represents the entropy.

The quantity $Q^+ \propto -d(R^2 \Omega^2)/dR$ represents the local energy generation rate, which serves as the energy source injected into the disk when material accretes inward. Conversely, $Q^-$ describes the radiative energy loss and is directly related to the local temperature $T$ \citep{Narayan1994}. When $Q^+$ becomes sufficiently large, the majority of the generated energy is channeled into driving turbulence rather than increasing the local temperature. This mechanism ensures the disk maintains a stable temperature and prevents radiative instability.

In thin disk region, turbulence is small, then $c_s \sim v_t < v_K$. As turbulence develops,  $c_s < v_t \sim v_K$. 
\rev{
Equations~(3)--(6) essentially decompose the sound speed squared $c_s^2$ in Equations~(1)--(4) of \citet{Narayan1994} into the sum of the thermal sound speed squared and the turbulent velocity squared, $c_s^2 + v_t^2$, and partition the internal energy term $T\,{\rm d}s$ into $T\,{\rm d}s + \frac{1}{2}\,{\rm d}v_t^2$. Additionally, they reformulate the energy input term $Q^+ \propto -{\rm d}(R^2\Omega^2)/{\rm d}R$ as a radius-dependent expression.
So it is expected that
} 
the \rev{scaling} solutions derived by Narayan and Yi \citeyearpar{Narayan1994} remain applicable for equations (3)--(6). Under these conditions, the physical parameters exhibit the following radial scaling relations:
$\rho \propto R^{-3/2}, \quad v_R \propto R^{-1/2}, \quad \Omega \propto R^{-3/2}, \quad c_s^2 \propto R^{-1}, \quad v_t^2 \propto R^{-1}$.

\section{Thick disk and jet formation}

The accretion material may form a standard thin disk at large radius $R$ \citep{Shakura1973}. As the material migrates inward, rising temperatures and increasing turbulence contribute to a substantial rise in total pressure, likely resulting in the formation of a thick disk structure.  
\rev{
According to \citet{Frank2002}, for a steady thick accretion disk in pure rotation about its rotation axis, the effective potential of a rotating blob in cylindrical coordinates is (\S10 in \citealt{Frank2002}; see also \citealt{Lynden-Bell1978, Abramowicz1984}):
\begin{equation}
\Phi_{\text{eff}} = -\frac{GM}{(R^2 + z^2)^{1/2}} + \frac{l^2}{2R^2}
\end{equation}
where $M$ is the mass of the central black hole, $R$ is the radius, $z$ is the height, and $l^2 = GMR_K$ is the square of the specific angular momentum at the Keplerian radius $R_K$. This equation can be rewritten as
\begin{equation}
e = \frac{2R_K\Phi_{\text{eff}}}{GM} = \left(\frac{R_K}{R}\right)^2 - \frac{2R_K}{\sqrt{R^2 + z^2}}
\end{equation}
where $e$ is the specific energy of the fluid in units of the binding energy of a Keplerian circular orbit of radius $R_K$, ranging between $-1$ and $0$.
Fig.~1. shows the effective equipotential field (see also Fig.~10.5 in \citealt{Frank2002}). The value $e=-1$ corresponds to pure Keplerian circular rotation at $R_K$, while other non-zero values of $e$ represent equipotential or total pressure contours. In the steady disk, the pressure contour can also be considered as the mass density profile, so Fig. 1. also represents the geometry of the thick disk cross section near the black hole. The ``funnel'' is formed above and below the central object at position $(0,0)$. When $e=0$, the contour is not closed, which means mass can flow away from the object. For $e>0$, the contour appears as two streams above and below the central object, which are identified as ``jets'' \citep{Frank2002}.
}

\rev{
\begin{figure}
  \includegraphics[width=\columnwidth]{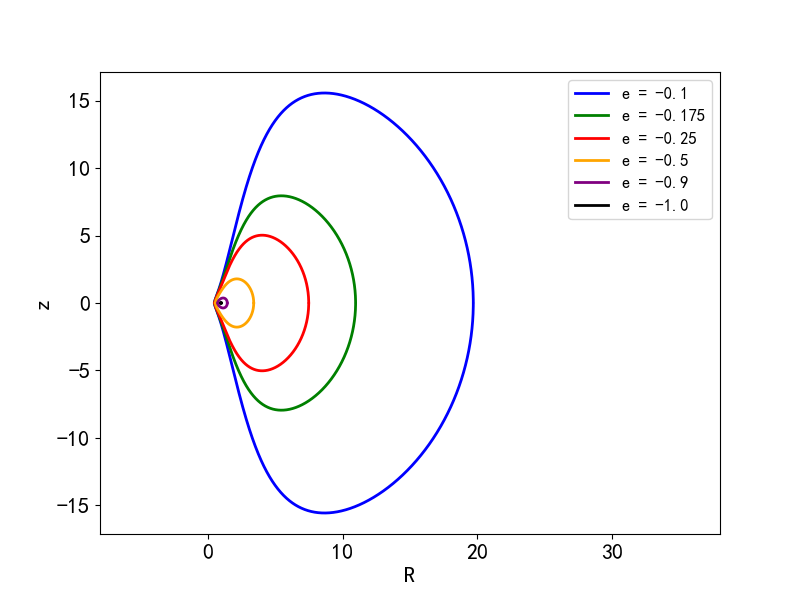}
 \caption{The equipotential field or the cross section shape of the thick disk for different equipotential energies. $e=-1$ stands for pure Keplerian orbital motions while other $e$ values stand for effective equipotential whose "energy" is larger than Keplerian one. Coordinates are in arbitrary scales.}
 \label{fig:jet_field}
\end{figure}
}

\rev{
The proposed ADAF disk is turbulent, so its shape might differ slightly from that of a steady thick disk; in particular, the relationship between total pressure and density may differ from that between thermal pressure and density, but the overall geometry should be similar. One comment should be added regarding the steady thick disk. The effective energy, $-1 \leq e < 0$, determines the shape and thickness of the disk. The excess energy above $-1$ originates from the pressure of the thick disk. The thicker the disk, the higher the pressure at the mid-plane, the higher the total disk energy ($e$ closer to $0$), and the more likely outflows and jets will form. This excess energy (above Keplerian) can be understood in a different way. The effective energy is calculated based on the conservation of angular momentum (Equation~7). Considering the motion of a gas blob, the excess energy arises from the difference between the actual velocity and the Keplerian velocity at that position. This excess energy causes the blob's orbit to cover other regions on the contour. In other words, different energies (generally exceeding Keplerian) cause the blobs to fill regions within the contour, thereby forming a thick disk. This concept will be used later to illustrate how turbulence can establish a thick disk and how the disk thickness relates to the turbulent energy.
}

A thick disk must maintain vertical pressure balance against the vertical component of the central object's gravity. 
Accordingly, we expect the midplane pressure to increase as 
material flows inward, implying $dP_{\text{tot}}/dR < 0$ \rev{(see Fig. 1)}. This pressure gradient creates an outward 
force that partially balances the gravity, leading us to 
expect a lower angular velocity than the Keplerian value, i.e., $\Omega<\Omega_K$. This result is also consistent with Narayan and Yi \citeyearpar{Narayan1994}. 
\del{
The balance among the gravity, centrifugal force, 
and pressure
gradients (both radial and vertical), together with the 
conservation of local angular momentum, can establish an 
equipotential field that molds the infalling material into a 
thick disk or torus with a central funnel 
In such a thick disk, the radial profile of the midplane pressure reaches a peak and then drops sharply near the center.
}
\rev{The radial profile of the midplane pressure reaches a peak and then drops sharply near the center.}
As a result, in the 
vicinity of the central object, the total pressure gradient 
satisfies $dP_{\text{tot}}/dR >0$ \rev{(see Fig. 1)}. This alignment of the 
pressure force with gravity enhances the net inward pull, 
causing the rotational velocity to exceed the Keplerian value.
Such super-Keplerian motion may serve as the fundamental driver of jet formation.

Let's examine the region where $dP_{\text{tot}}/dR > 0$ in more detail. Set $GM = 1$. For Keplerian rotation, the gravitational potential energy is $\Psi = -1/R$, the kinetic energy (or local Keplerian energy) is $E_k = 1/(2R)$, the total energy is $E = \Psi + E_k = -1/(2R)$, and the angular momentum is $L = R^{1/2}$.

Consider a blob being pushed inward by pressure. Let the initial and final positions be $R_i$ and $R_f$ with $R_i/R_f > 1$. With conservation of angular momentum, the final speed becomes $R_i/R_f$ times the initial speed, and the final kinetic energy becomes $(R_i/R_f)^2$ times the initial kinetic energy. The total energy of the blob increases by an amount
$\Delta E = \left( R_i / R_f -1 \right)^2 /(2 R_i)$.
The blob gains energy because pressure does work on it. This process is analogous to a ball on a rope being swung in a circle: when you pull the rope inward to make the circle smaller, the ball speeds up to conserve angular momentum. A large ratio $R_i/R_f$ significantly increases the energy gain $\Delta E$. 
\rev{
If the pressure gradient acts on a static blob, it will only accelerate the blob toward the central object. When the pressure gradient acts on a rotating blob (not necessarily in thin-disk Keplerian motion, as the direction can be random due to turbulence), it will accelerate the blob along its direction of motion rather than pushing it toward the center. Since the direction of motion of a turbulent blob is random, some blobs, after being accelerated, may point outward along the funnel rather than toward the central object. 
} 
\del{Even for a small ratio $R_i/R_f$, if the process occurs step by step, the blob may still accumulate a large amount of energy.}

\rev{
The energy gain $\Delta E$ may not reach $\left(R_i/R_f - 1\right)^2/(2R_i)$ when $R_i/R_f > 2$. Since the energy imparted to the blob ultimately originates from the pressure difference. The pressure difference is always less than $E_K$ of $R_i$ between $R_i$ and envelop; otherwise, the total energy of the disk would exceed zero and the disk would overflow to infinity (see Fig. 1.). If the blob's initial kinetic energy is $E_K$, then its final kinetic energy will be less than $2E_K$, which is smaller than the escape velocity.
}
\rev{
In fact, there is a innermost position $R_{in}$ for a blob at $R_i$ to reach \citep{Frank2002}
\begin{equation}
R_{in} = \frac{R_i}{1 + \sqrt{1 + e}} = \frac{R_i}{1 + \sqrt{\eta}}
\end{equation}
where $\eta=1+e$  is a value will be defined in next paragraph. Take $R_{in}$ as $R_f$, then $R_i/R_f = 1+\sqrt{1 + e}$ is always smaller than 2, which is consistent with the requirement $R_i/R_f < 2$. The time scale for the acceleration is of the order of orbital period, since the random speed of the blob is of the order of Keplerian speed.
}

\del{Suppose $R_f$ is the inner edge of the torus or the starting point of the funnel. If the blob's kinetic energy exceeds $3E_k (R_f)$, with two-thirds of this kinetic energy balancing the gravitational binding energy, the blob may escape from the central object through the funnel at Keplerian speed. The angular momentum of a blob that has kinetic energy $3E_k(R_f)$ at radius $R_f$ originates from a blob with kinetic energy $E_k(R_i)$ at a radius three times larger, i.e., $R_i = 3R_f$. This implies that the acceleration of such blobs occurs in a region very close to the central object. Since the total energy and angular momentum of the disk are conserved, the energy transfer mechanism described above merely redistributes these quantities among different blobs, leaving some blobs more energized than others.}

\del{With the fundamentals in place—namely, the formation of thick disks, tori, and acceleration mechanisms—we now turn to a crucial ingredient that ties these processes together: turbulence.} The differential rotation of the disk generates local turbulence through Rayleigh–Taylor and Kelvin–Helmholtz instabilities \citep{Frank2002}. The well-known $\alpha$-prescription introduced by Shakura and Sunyaev \citeyearpar{Shakura1973} is generally attributed to the turbulence. Near the central source in the torus, instabilities become violent and fully developed turbulence is established. Blobs of various sizes emerge, with the smallest stable structures—characterized by low Reynolds numbers and nearly isotropic velocities—surviving at the viscous scale. These small-scale blobs or eddies are embedded within, and strained by, larger structures; their dynamics remain coupled to the large-scale flow, with energy spectra following Kolmogorov's scaling \citep{Pope2000}. 
\rev{
Assume that the turbulent energy in the disk at radius $R_K$ is a fraction $\eta$ of the local Keplerian energy $E_K$, i.e., $E_t = \eta E_K$; then the velocity dispersion $\sigma$ of the turbulence relative to the Keplerian speed is $\sigma^2 = \frac{2}{3} E_t = \frac{2}{3} \eta E_K$ (\S12 in \citealt{Pope2000}). The relative velocity distribution of the turbulence in the ADAF disk is then:
\begin{equation}
f(\mathbf{v} - \mathbf{v}_K) = \frac{1}{(2\pi)^{3/2} \sigma^3} \exp\left( -\frac{v_R^2 + v_z^2 + \delta v_\phi^2}{2\sigma^2} \right)
\end{equation}
where $v_R$ and $v_z$ are the $R$ and $z$ velocity components, and $\delta v_\phi$ is the rotational velocity component relative to the Keplerian value $v_\phi$. The velocity distribution function can also be interpreted as the distribution of velocities of the gas blobs. We note that the value $\eta$ and the effective energy $e$ in Equation~(8) have a simple relationship: $e = \eta - 1$, i.e., the total energy in the disk is the sum of the turbulent and Keplerian energies minus the gravitational energy. Also, the height of the thick disk $H$ is proportional (though not linearly; see Fig. 1.) to the parameter $\eta$. In accretion disks, the random velocity may be directionally dependent, and the distribution may not be Gaussian \citep{Pan2014,Stoll2016}. Here we adopt the Gaussian distribution function only for a simple estimate. From Equation~(10), there must be some blobs whose velocities exceed $1\sigma$ (the probability is about $3.1\%$ for all velocity components to exceed $1\sigma$); thus their initial kinetic energy exceeds $(1+\eta)E_K$ (note that we omit the cross term between $v_\phi$ and $\delta v_\phi$ for simplicity in calculation). Adding the energy $\eta E_K$ gained from pressure acceleration, the total kinetic energy can exceed $(1+2\eta)E_K$. When $\eta > 0.5$, a blob may gain total kinetic energy exceeding the escape energy $2E_K$. If $\eta \approx 1$, then some blobs' final energy will exceed $3E_K$.
}

\rev{
The previous discussion applies at radius $R_K$, but it is also applicable to other radii except at the innermost radius where the funnel is formed. At this boundary of the disk, the estimation might deviate more significantly. In general, however, the turbulent energy and the acceleration by pressure can enable part of the turbulent blobs to reach $2\sim3$ times the local Keplerian energy $E_K$.
}

\rev{With the energy redistribution by turbulence and the final acceleration by pressure gradient,}
\del{Through this acceleration mechanism and Kolmogorov scaling,} 
some blobs may attain energies exceeding $3E_K$ and be directed outward along the funnel, escaping the central object at roughly Keplerian speed. It is expected that twin jets form in opposite directions along the funnels, and that the two jets may be statistically similar. Statistically, this also implies that the jet speed and mass flux may vary over time. Blobs with energies below $2E_K$ may fall back onto the central object or the torus. Blobs not directed toward the funnel opening may either accrete onto the central object or return to the torus for another \rev{turbulence and } acceleration cycle. As discussed earlier, only acceleration occurring in the very central region \del{—within about three radii of the central object—} is likely to contribute to jet formation. In other regions of the torus, this process merely redistributes momentum and energy among the rotating material.

The structure and opening angle of the funnel are likely more complex than the steady-state configurations presented in \rev{Fig. 1. or} Lynden-Bell \citeyearpar{Lynden-Bell1978} or Frank \citeyearpar{Frank2002}, largely due to the influence of turbulence. Nevertheless, the overall geometry is expected to remain qualitatively similar. The degree of jet collimation is primarily governed by the funnel's opening angles. Near the midplane, the funnel and the central spherical object define a ring-like region where the most powerful jet blobs are generated. This suggests that the jet may consist of concentric cylindrical layers around the central axis. Blobs originating from the outer regions of the funnels—located slightly farther from the central object—are likely to produce broader and slower outflows.

Now let us revisit the formation conditions for an advection-dominated accretion flow. The net energy stored in the accretion flow consists of two components: thermal energy and turbulent energy. There exists a characteristic timescale known as the dissipation time, denoted as $t_d \approx (3-5) \, l/v_{\text{rms}}$, where $l$ represents the largest turbulent eddy scale and $v_{\text{rms}}$ is the one-dimensional root-mean-square velocity associated with that scale, $(3-5)$ are typical eddy turnover time  \citep{Pope2000,Elmegreen_1985}. $v_{\text{rms}}$ is related to the three-dimensional turbulent velocity $v_t$ as $v_t^2 = 3v_{\text{rms}}^2$ 
\rev{(note: $v_{\text{rms}} = \sigma$ in Eq. (10))}. 
We can use $2H$ to replace $l$, thus $t_d \approx (6-10) \sqrt{3} \, H/v_t$. It is understandable that when the disk becomes thick, the size of the largest eddy $2H$ increases, leading to a longer dissipation time and making it much easier for the turbulence to retain energy. After this timescale, the turbulent energy is converted into thermal energy.

The energy input rate $Q^+$ can be understood from an alternative perspective. Consider a ring of accreting material that moves inward by a radial distance $\Delta R$. The gravitational binding energy released in this process is entirely converted into turbulent energy. The radial dynamical timescale is defined as $t_R \approx \Delta R / v_R$, where $v_R$ is the radial inflow velocity. If the turbulent dissipation timescale $t_d$ is shorter than the radial drift timescale ($t_d < t_R$), turbulence does not have sufficient time to grow significantly, and the accretion disk remains geometrically thin. Conversely, if $t_d > t_R$, a larger fraction of the released energy can be stored in turbulent motions, leading to a thickened disk and the onset of an ADAF.

According to the standard thin disk model, the radial velocity scales as $v_R \approx \alpha c_s (H/R)$, and the turbulent velocity is given by $v_t = \alpha c_s$ \citep{Frank2002}. Replacing the radial step size $\Delta R$ with the disk scale height $H$, the radial timescale becomes $t_R \approx \Delta R / v_R = H / v_R = R / v_t$. Assuming that the transition to an ADAF occurs when the dissipation timescale is comparable to the radial drift timescale ($t_d \approx t_R$), we obtain
$R/H \approx (6-10) \times \sqrt{3} \approx 15$.
We may roughly take $H$ as the innermost radius of the disk and estimate that the ADAF starts around 15 times of this radius. This estimation is very rough and can be off by order of magnitude though.

The stability of thick accretion disks has long been a subject of concern, with thermal and hydrodynamic instabilities extensively discussed in the literature \citep{Papaloizou1984,Abramowicz1988}. Local instabilities are generally expected to contribute only to turbulent activity and are not sufficient to disrupt the overall accretion flow. In contrast, global instabilities such as PPI may pose more significant challenges. The PPI was originally analyzed in the context of a static, rotating torus using linear perturbation theory. However, whether this instability remains problematic in a torus that is already turbulent is not immediately clear. The global conservation of total energy and angular momentum imposes constraints on the large-scale structure of thick disks; instabilities may therefore only enhance turbulence within an already chaotic flow without fundamentally altering its configuration. Supporting this view, Hawley \citeyearpar{Hawley1991} demonstrated through numerical simulations that the unstable modes can saturate at low amplitudes, suggesting that non-axisymmetric instabilities are not sufficiently robust to preclude thick tori from serving as viable models for accretion disks.

\rev{
As a brief summary of the model described in \S2 and \S3, we list the following results here: (1) The turbulence, which stores the binding energy from the accretion disk, helps to establish the ADAF thick disk, which is an important condition for jet formation; (2) The turbulence redistributes the flow energy and velocity, and the thick ADAF establishes the positive pressure gradient to accelerate gas blobs; both mechanisms combine to enable some gas blobs to exceed the escape velocity. The ADAF itself does not create the jet; rather, the velocity dispersion in turbulence and the acceleration of the blobs produces the jet; (3) $\eta$, which indicates the fraction of Keplerian energy stored in turbulence, is a crucial parameter that determines whether a jet can form. $\eta > 0.5$ is the minimal value required to initiate the jet. $\eta$ also determines the height and size of the thick disk and the opening angle of the disk (see Fig. 1.); (4) The \revs{proposed turbulent} ADAF disk is generally optically thick, as it represents a simple geometric extension of the thin disk, in contrast to the original ADAF disk in \citet{Narayan1994}, which is optically thin; (5) The jet originates in the innermost small region near the center of the funnel. For AGN, this region is expected to be near the innermost stable circular orbit (ISCO); for YSOs, it is near the surface of the star\revs{; (6) The majority of the jet content is assumed to be normal (baryonic) material similar to that in the accretion disk rather than electron-positron pairs in black hole systems}. 
} 

\section{Application to different astronomical objects}
The nearby supermassive black hole M87, with a mass of $M = 6.5 \times 10^{9} \, M_{\odot}$ (where $M_{\odot}$ is the solar mass), exhibits giant radio jets extending to kiloparsec scales \citep{Owen1989,Biretta1999}. This source was recently observed using the Event Horizon Telescope (EHT) at $1.3~\mathrm{mm}$ and the Global Millimeter VLBI Array (GMVA) at $3.5~\mathrm{mm}$ \citep{Lu2023}. High-resolution images reveal a ring-like structure with diameters of approximately $5.5 \, R_{\mathrm{s}}$ (where $R_{\mathrm{s}}$ is the Schwarzschild radius) at $1.3~\mathrm{mm}$ and $8.4 \, R_{\mathrm{s}}$ at $3.5~\mathrm{mm}$, interpreted as part of the accretion flow. The inner edge of the flow may even lie inside the innermost stable circular orbit at $3 \, R_{\mathrm{s}}$. The images also show an edge-brightened jet connecting to the black hole's accretion flow. These observational features are consistent with our proposed model, in which jets consist of blobs ejected from the innermost region of a thick accretion disk. The sub-parsec radio jets on scales within $100 \, R_{\mathrm{s}}$ display triple-ridge structures, which can be understood as cylindrical layers as the model expects. 
Punsly \citeyearpar{Punsly2024}, in a study of the M87 jets, reached a similar conclusion: the axial jet appears to emerge from a nozzle in a thick accretion disk.
\rev{
\citet{Lu2023} concluded that the observations were consistent with the \citet{Blandford1977} mechanism and ruled out the magneto-centrifugal force related disk-jet model based on the limb-brightened effects seen in M87 jets. \citet{Yang2024} performed GRMHD simulations and reached similar results. However, whether the Blandford-Znajek mechanism really works the way it was originally proposed is still questionable. \citet{King2021} found that the Blandford-Znajek mechanism cannot tap the spin energy of a black hole continuously and is not a viable mechanism for powering continuous astrophysical jets. They also suggest that the jet should be powered by the accretion disk rather than the central black hole, as proposed in this work.
}

In a review article, Yuan and Narayan \citeyearpar{Yuan2014} show that AGN jets are more likely to originate from thick accretion disks rather than thin ones. They also point out that in many low-luminosity AGNs, the accretion flow likely consists of an outer thin disk that transitions into a hot, thick flow in the inner region. This configuration is consistent with our model, where turbulence dominates in the inner region and leads to the formation of a thick disk.

At the lower end of the central mass scale, young stellar objects with masses ranging from approximately $0.1$ to $100 M_{\odot}$ display jet structures that bear a remarkable resemblance to those observed in active galactic nuclei. Using ALMA, \citet{Lee2023} detected in the YSO object HH211 both a disk, traced by H$_2$ continuum emission at 352~GHz, and inner jets observed in molecular lines of CO and SiO \citep{Lee2025}. The jets are oriented nearly perpendicular to the disk and display coaxial layering, with the CO jet confined within the central axis of the SiO jet. Measurements of specific angular momentum suggest that the jet is launched from the innermost edge of the disk. All these observed features are consistent with the model proposed in this paper.

The brightness temperatures of the CO and SiO jets are below 50~K, while the disk temperature is around 40~K \citep{Lee2023, Lee2025}, indicating that the materials in both the disk and jets are largely unionized. This poses a challenge for mechanisms that rely on magneto-centrifugal forces to accelerate neutral gas to Keplerian speeds from a rotating disk. Even if the inner disk region were ionized by the central star, it remains difficult to explain how the jet could cool so rapidly back to $\sim$50~K. 

Notably, jet formation is far more common in YSOs: virtually all low- and high-mass stars produce accretion-powered, collimated outflows during their formation \citep{Frank2014}. In contrast, only about $10\%$ of AGNs are classified as jet-producing, or radio-loud \citep{Kellermann1989, Xu1999}. This disparity can be understood in terms of disk temperature. In YSOs, the accretion disks are very cool, and because radiative energy loss scales as $T^4$, most of the energy generated in the disk remains trapped, likely sustaining turbulence. This may naturally lead to thicker disks and more frequent jet production in the YSO context.

X-ray binaries represent a class of systems that frequently exhibit accretion disks and associated jets. In a comprehensive review of X-ray observations of accretion disks, \citet{Inoue2022} proposed a direct correspondence between the primary spectral states of X-ray binaries---namely the high-soft state, the low-hard state, and the very high state---and distinct accretion disk configurations: the standard thin disk, the advection-dominated accretion flow, and the slim disk, respectively. Notably, jets are observed exclusively during the low-hard state and are absent in the high-soft state, suggesting that jet formation is closely tied to the presence of a thick (ADAF-type) accretion flow---a condition consistent with the proposed models of jet launching in this paper.

\rev{
In a subsequent paper \citep{Xu2026}, we proposed that the ADAF disk is variable and that a complete cycle of ADAF contraction, transition to a thin disk, and subsequent re-expansion corresponds to the rapid rise, peak, and decay phases observed in the X-ray outbursts of black hole XRBs. With the introduction of the parameter $\eta$, it is clear that the size of the ADAF disk is governed by this parameter. During the decay period of the X-ray outburst, $\eta$ increases, and the size of the ADAF grows accordingly. The XRB then enters the low-hard state. When $\eta > 0.5$, the jet \revs{may form}\dels{forms}, consistent with observations \citep{Inoue2022,belloni2010}. The jet is absent in the high-soft state because the ADAF disk is small and $\eta < 0.5$.
}

The disk configurations \rev{in black hole XRBs} associated with different X-ray states \citep{Inoue2022,Fender2004} bear strong resemblance to those described in \citeauthor{Yuan2014}'s \citeyearpar{Yuan2014} review of active galactic nuclei, where analogous accretion disk modes correspond to various classes of AGNs. This parallelism points to a common underlying physical framework \rev{, i.e., the black hole and the thick disk,} linking X-ray binaries and AGNs, despite their differences in scale and environment.

Jets in X-ray binaries are predominantly observed in black hole systems \citep{Fender2001,Fender2004} and are comparatively rare in neutron star binaries. The most plausible explanation is that neutron stars often possess strong magnetic fields, which can disrupt the inner accretion disk when it approaches too closely. In such cases, the formation of a thick inner disk is inhibited, thereby suppressing jet production. The characteristic radius at which the magnetic field begins to significantly influence the accretion flow is typically on the order of $10^7$--$10^8$~cm, substantially larger than the neutron star radius of about $10^6$~cm \citep{Spruit1993} which corresponds to the innermost radius of the funnel for jet to form. Under these considerations, we are inclined to believe that the binary systems containing powerful jets---such as SS~433 and Cygnus~X-3, as listed in Table 1 of Mirabel and Rodríguez \citeyearpar{Mirabel1999} as \rev{neutron star} --- are black holes rather than neutron stars.

\revs{
\section{Caveats and Future Work}
The present model is inherently partially phenomenological. It is built upon a set of steady-state related assumptions that simplify the complex physics involved. In reality, the governing equations are turbulent and highly nonlinear, and our steady-state treatment inevitably smooths over time-dependent behavior that may be critical to the full dynamics. Consequently, the current framework does not constitute a solution directly derived from the fundamental equations, but rather a physically motivated approximation that captures certain aspects of the system, i.e., the formation of thick disk, funnels, the energy budget of the jet and the location of the jet base. 
}

\revs{
Several specific physical effects have been omitted. For instance, we have not considered the back-reaction (dynamical) effect of the accelerating blob on the surrounding medium, nor the influence of random or uneven jet structures on the turbulent ADAF. The potential role of a spinning black hole with dynamic spacetime — including frame-dragging or other mechanisms — is also not included. These omissions represent important limitations of the current model.
}

\revs{
On the observational side, our comparison with observations remains qualitative rather than quantitative. The model in its present form cannot constrain key physical quantities such as the jet flux, total energy, or detailed spectra. While this level of comparison suffices to demonstrate qualitative consistency, a more rigorous quantitative test against observations will require further development.
}

\revs{
Future work should aim to overcome these limitations. A fully global analytical solution, while challenging, would provide a firmer theoretical foundation. More realistically, numerical simulations that relax the steady-state and axisymmetric assumptions, incorporate turbulence and nonlinearity, and include the missing dynamical effects (back-reaction, uneven jets, spinning black hole spacetime) are urgently needed. Such efforts will be essential to determine whether the proposed mechanism can operate self-consistently in a full global outflow solution, and to enable quantitative predictions that can be directly compared with observations.
}

\section{Summary and Discussions}

Astrophysical jets are discovered in many classes of astronomical objects, including AGN, YSO, XRB and possibly some other objects. The jets are well collimated, at least at the starting points of the jets, and their velocities are close to the escape velocity of the central objects. A variety of models have been developed to explain the formation of the jets. Some models rely on the energy of the central objects, others rely on the rotating energy of the disks, but the exact mechanism to fully support the jet formation and collimation is still not well developed. We believe, as also discussed by \cite{Livio1999}, that the mechanism of jet formation in different classes of objects MUST be the same. The prevailing magnetocentrifugal model for jet formation may not apply universally across diverse astrophysical systems. Accretion disks around protostars and black holes—whether spinning or non-spinning, stellar-mass or supermassive—exhibit vastly different mass densities, temperatures, and \rev{especially the} ionization fractions. And the origin of the requisite large-scale magnetic fields remains uncertain. It is therefore difficult to envision a single magnetocentrifugal mechanism producing similar jets in such disparate environments. This concern motivates us to seek jet formation mechanisms under more fundamental conditions—specifically, at the pure hydrodynamic level.

The key innovation we introduce to existing ADAF models is turbulence. Traditional ADAF formation requires low radiation efficiency, typically attributed to the ion-electron two-temperature assumption. While this two-temperature plasma may apply to AGN or X-ray binary accretion disks, it appears unsuitable for YSO disks, where temperatures remain below $\sim$200 K \citep{Ueda2023, Lee2025}, rendering the two-temperature model irrelevant for YSO systems.

Turbulence, however, can store much of the gravitational binding energy released in the disk and is naturally expected in rotating systems, whether AGN or YSO disks. This turbulent energy eventually thermalizes after a dissipation time $t_d$. When $t_d$ exceeds the energy loading time $t_r$, an ADAF forms. In this framework, the two-temperature plasma is unnecessary, instead, the released energy is temporarily stored in turbulence, enabling ADAF formation. Consequently, our \revs{turbulent} ADAF model applies automatically to both AGN and YSO, or other similar systems. As the reservoir for most of the binding energy, the advection-dominated accretion flow serves as the energy source that powers the outflows.

Turbulence also generates clusters of the smallest stable fluid elements, each with distinct velocities and directions. \rev{The speed of the fluid blobs follow a Gaussian-like distribution, with some blobs carrying higher than Keplerian energy}. These structures can persist over extended periods and may be further accelerated \rev{by pressure gradient with conservation of angular momentum.} \del{through the mechanism discussed earlier.} The formation of jets require a thick accretion disk \rev{within which a fraction $\eta  ( > 0.5) $ of local Keplerian energy is stored in the turbulence, and} with funnel-like regions to confine and direct the escaping blobs.
\rev{The key results of the ADAF disk and jet formation mechanism are summarized at the end of \S3 .}
The jet base is expected to form within the innermost funnel near the central object. This scenario is \rev{consistent with} \del{supported by} numerous observations across a range of systems, from active galactic nuclei to young stellar objects. We believe that jet formation is fundamentally hydrodynamic in nature, with magnetic fields and radiation playing only supporting roles.

This model integrates the accretion disk, advection-dominated flow, thick inner disk, acceleration mechanism, jet formation and collimation scheme into a unified framework applicable to AGN, YSO, XRB, and other similar systems. We estimate that \del{the acceleration zone for the jet blobs lies within approximately three times the inner disk radius, while} the ADAF begins \rev{around} \del{beyond} 15 times the inner disk radius \rev{but is really dependent on the value of $\eta$}. The jet is formed through the accumulation of high-speed turbulent blobs, rather than emerging directly from global solutions of the ADAF equations. The mass flux, energy, and jet velocity are difficult to constrain \rev{within the scope of this work}, as \rev{turbulences} \del{they} appear to be dynamic. \revs{This is definitely a strong limitation of this work - a partially phenomenological model.} While computer simulations suggest that ADAF promotes outflow formation in the polar regions \citep{Stone1999, Yuan2012}, these outflows differ significantly from collimated jets. This discrepancy may arise because the simulation grids are too coarse to resolve the smallest turbulent blobs. We propose that higher-resolution simulations capable of capturing full turbulence could better predict the mass, speed, and energy of jets.

\section*{Acknowledgements}

\rev{The author is grateful to the anonymous referee for the careful and insightful review. Their comments have substantially strengthened the rigor and clarity of this work.} The author \rev{also} wishes to express sincere gratitude to Dr. Chin-Fei Lee for sharing his latest observational results on HH211, which have rekindled my interest in the jet formation mechanism. 
This work is \del{partially} supported by the China Manned Space Program \rev{Grant No. CMS-CSST-2025-A19 and Grants } \del{through grants} allocated for the development of the Multi-Channel Imager and the Integral Field Spectrograph, led by Drs. Zhenya Zheng and Lei Hao respectively, for the Chinese Space Station Telescope (CSST).

\section*{Data Availability}

No new data were created or analysed in this study.




\bibliographystyle{mnras}
\bibliography{example} 

@article{Abramowicz1984,
    author = {Abramowicz, M. A. and Livio, M. and Piran, T. and Wiita, P. J.},
    title = {Local stability of thick accretion disks. I - Basic equations and parallel perturbations in the negligible viscosity case},
    journal = {\apj},
    year = {1984},
    volume = {279},
    pages = {367},
    doi = {10.1086/161898}
}

@article{Abramowicz1980,
    author = {Abramowicz, M. A. and Piran, T.},
    title = {On collimation of relativistic jets from quasars},
    journal = {\apj},
    year = {1980},
    volume = {241},
    pages = {L7-L11},
    doi = {10.1086/183349}
}

@article{Abramowicz1995,
    author = {Abramowicz, Marek A. and Chen, Xingming and 
    Kato, Shoji and Lasota, Jean-Pierre and Regev, Oded},
    title = {Thermal Equilibria of Accretion Disks},
    journal = {\apj},
    year = {1995},
    volume = {438},
    pages = {L37},
    doi = {10.1086/187709},
    archiveprefix = {arXiv},
    eprint = {astro-ph/9409018}
}

@article{Abramowicz1988,
    author = {Abramowicz, M. A. and Czerny, B. and Lasota,
    J. P. and Szuszkiewicz, E.},
    title = {Slim accretion disks},
    journal = {\apj},
    year = {1988},
    volume = {332},
    pages = {646},
    doi = {10.1086/166683}
}

@article{Anderson2005,
    author = {Anderson, J. M. and Li, Zhi-Yun and Krasnopolsky, R. and Blandford, R.D.},
    title = {The Structure of Magnetocentrifugal Winds. I. Steady Mass Loading},
    journal = {\apj},
    year = {2005},
    volume = {630},
    pages = {945},
    doi = {10.1086/432040}
}

@article{Blandford1999,
    author = {Blandford, R. D. and Begelman, M. C.},
    title = {On the fate of gas accreting at a low rate on to a black hole},
    journal = {\mnras},
    volume = {303},
    number = {1},
    pages = {L1--L5},
    year = {1999},
    doi = {10.1046/j.1365-8711.1999.02358.x}
}

@article{Blandford1982,
    author = {Blandford, R. D. and Payne, D. G.},
    title = {Hydromagnetic flows from accretion discs and the production of radio jets},
    journal = {\mnras},
    volume = {199},
    number = {4},
    pages = {883--903},
    year = {1982},
    doi = {10.1093/mnras/199.4.883}
}

@article{Blandford1977,
    author = {Blandford, R. D. and Znajek, R. L.},
    title = {Electromagnetic extraction of energy from Kerr black holes},
    journal = {\mnras},
    volume = {179},
    number = {3},
    pages = {433--456},
    year = {1977},
    doi = {10.1093/mnras/179.3.433}
}

@article{Biretta1999,
    author = {Biretta, J. A. and Sparks, W. B. and Macchetto, F.},
    title = {Hubble Space Telescope Observations of Superluminal Motion in the M87 Jet},
    journal = {\apj},
    volume = {520},
    number = {2},
    pages = {621--626},
    year = {1999},
    doi = {10.1086/307499}
}

@incollection{Frank2014,
    author = {Frank, A. and others},
    title = {Jets and Outflows from Star to Cloud: Observations Confront Theory},
    booktitle = {Protostars and Planets VI},
    pages = {451--474},
    publisher = {The University of Arizona Press},
    address = {Tucson},
    year = {2014},
    doi={10.48550/arXiv.1402.3553}
}

@book{Frank2002,
    author = {Frank, J. and King, A. and Raine, D. J.},
    title = {Accretion Power in Astrophysics},
    publisher = {Cambridge University Press},
    address = {Cambridge, UK},
    year = {2002},
    isbn = {978-0521629577},
    doi = {10.1017/CBO9781139164245}
}

@article{Fender2001,
       author = {{Fender}, R.~P.},
        title = "{Powerful jets from black hole X-ray binaries in low/hard X-ray states}",
      journal = {\mnras},
         year = 2001,
       volume = {322},
        pages = {31-42},
          doi = {10.1046/j.1365-8711.2001.04080.x},
}

@article{Fender2004,
    author = {Fender, R. P. and Belloni, T. M. and Gallo, E.},
    title = {Towards a unified model for black hole X-ray binary jets},
    journal = {MNRAS},
    volume = {355},
    pages = {1105--1118},
    year = {2004},
    doi = {10.1111/j.1365-2966.2004.08384.x}
}

@article{Hawley1991,
    author = {Hawley, J. F.},
    title = {Three-dimensional simulations of black hole tori},
    journal = {ApJ},
    volume = {381},
    pages = {496--507},
    year = {1991},
    doi = {10.1086/170673}
}

@article{Inoue2022,
    author = {Inoue, H.},
    title = {X-ray observations of accretion disks},
    journal = {\pasj},
    volume = {74},
    pages = {1},
    year = {2022},
    doi = {10.1093/pasj/psab066}
}

@article{Kellermann1989,
    author = {Kellermann, K. I. and Sramek, R. and Schmidt, M. and Shaffer, D. B. and Green, R.},
    title = {VLA Observations of Objects in the Palomar Bright Quasar Survey},
    journal = {AJ},
    volume = {98},
    pages = {1195--1207},
    year = {1989},
    doi = {10.1086/115207}
}

@article{Lee2025,
    author = {Lee, C. F. and Jhan, K. S. and Moraghan, A.},
    title = {A magnetized protostellar jet launched from the innermost disk at the truncation radius},
    journal = {Scientific Reports},
    volume = {15},
    pages = {29702},
    year = {2025},
    doi = {10.1038/s41598-025-11602-w}
}

@article{Lee2023,
    author = {Lee, C. F. and Jhan, K. S. and Moraghan, A.},
    title = {First Detection of a Linear Structure in the Midplane of the Young HH 211 Protostellar Disk: A Spiral Arm?},
    journal = {ApJL},
    volume = {951},
    pages = {L2},
    year = {2023},
    doi = {10.3847/2041-8213/acdbca}
}

@article{Livio1999,
    author = {Livio, M.},
    title = {Astrophysical jets: a phenomenological examination of acceleration and collimation},
    journal = {Physics Reports},
    volume = {311},
    pages = {225--235},
    year = {1999},
    doi = {10.1016/S0370-1573(98)00102-1}
}

@article{Lu2023,
    author = {Lu, R. S. and Asada, K. and Krichbaum, T. P. and others},
    title = {A ring-like accretion structure in M87 connecting its black hole and jet},
    journal = {Nature},
    volume = {616},
    pages = {686--690},
    year = {2023},
    doi = {10.1038/s41586-023-05843-w}
}

@article{Lubow1994,
    author = {Lubow, S. H. and Papaloizou, J. C. B. and Pringle, J. E.},
    title = {Magnetic field dragging in accretion discs},
    journal = {\mnras},
    volume = {267},
    pages = {235},
    year = {1994},
    doi = {10.1093/mnras/267.2.235},
    adsurl={https://ui.adsabs.harvard.edu/abs/1994MNRAS.267..235L}
}

@article{Lynden-Bell1978,
    author = {Lynden-Bell, D.},
    title = {Gravity power},
    journal = {Physica Scripta},
    volume = {17},
    pages = {185--191},
    year = {1978},
    doi = {10.1088/0031-8949/17/3/009}
}

@article{Mirabel1999,
    author = {Mirabel, I. F. and Rodriguez, L. F.},
    title = {Sources of relativistic jets in the galaxy},
    journal = {ARAA},
    volume = {37},
    pages = {409--443},
    year = {1999},
    doi = {10.1146/annurev.astro.37.1.409}
}

@article{Narayan1994,
    author = {Narayan, R. and Yi, I.},
    title = {Advection-dominated accretion: A self-similar solution},
    journal = {ApJ},
    volume = {428},
    pages = {L13--L16},
    year = {1994},
    doi = {10.1086/187381}
}

@article{Narayan1995,
    author = {Narayan, R. and Yi, I.},
    title = {Advection-dominated accretion: Underfed black holes and neutron stars},
    journal = {ApJ},
    volume = {452},
    pages = {710--735},
    year = {1995},
    doi = {10.1086/176343}
}

@article{Owen1989,
    author = {Owen, F. N. and Hardee, P. E. and Cornwell, T. J.},
    title = {High-resolution VLA observations of M87},
    journal = {ApJ},
    volume = {340},
    pages = {698--707},
    year = {1989},
    doi = {10.1086/167430}
}

@article{Papaloizou1984,
    author = {Papaloizou, J. C. B. and Pringle, J. E.},
    title = {The dynamical stability of differentially rotating discs with constant specific angular momentum},
    journal = {MNRAS},
    volume = {208},
    pages = {721--750},
    year = {1984},
    doi = {10.1093/mnras/208.4.721}
}

@book{Pope2000,
    author = {Pope, S. B.},
    title = {Turbulent Flows},
    publisher = {Cambridge University Press},
    address = {Cambridge, UK},
    year = {2000},
    doi = {10.1017/CBO9781316179475}
}

@article{Punsly2024,
    author = {Punsly, B.},
    title = {First image of a jet launching from a black hole accretion system: Kinematics},
    journal = {\aap},
    volume = {685},
    pages = {L3},
    year = {2024},
    doi={10.1051/0004-6361/202449956}
}

@article{Shakura1973,
    author = {Shakura, N. I. and Sunyaev, R. A.},
    title = {Black holes in binary systems: Observational appearance},
    journal = {\aap},
    volume = {24},
    pages = {337--355},
    year = {1973},
    adsurl = {https://ui.adsabs.harvard.edu/abs/1973A&A....24..337S}
}

@article{Shu1994,
    author = {Shu, F. H. and Najita, J. and Ostriker, E. and Wilkin, F. and Ruden, S. and Lizano, S.},
    title = {Magnetocentrifugally driven flows from young stars and disks: A unified model},
    journal = {ApJ},
    volume = {429},
    pages = {781--796},
    year = {1994},
    doi = {10.1086/174363}
}

@article{Stone1999,
    author = {Stone, J. M. and Pringle, J. E. and Begelman, M. C.},
    title = {Hydrodynamical non-radiative accretion flows in two dimensions},
    journal = {MNRAS},
    volume = {310},
    pages = {1002--1016},
    year = {1999},
    doi = {10.1046/j.1365-8711.1999.03024.x}
}

@article{Spruit1993,
    author = {Spruit, H. C. and Tamm, R. E.},
    title = {An Instability Associated with a Magnetosphere-Disk Interaction},
    journal = {ApJ},
    volume = {402},
    pages = {593--603},
    year = {1993},
    doi = {10.1086/172162}
}

@article{Ueda2023,
    author = {Ueda, T. and Okuzumi, S. and Kataoka, A. and Flock, M.},
    title = {Probing the temperature structure of the inner region of a protoplanetary disk},
    journal = {\aap},
    volume = {675},
    pages = {A176},
    year = {2023},
    doi = {10.1051/0004-6361/202346253}
}

@article{Xu1999,
    author = {Xu, C. and Livio, M. and Baum, S.},
    title = {Radio-loud and Radio-quiet Active Galactic Nuclei},
    journal = {AJ},
    volume = {118},
    pages = {1169--1184},
    year = {1999},
    doi = {10.1086/301007}
}

@article{Yuan2014,
    author = {Yuan, F. and Narayan, R.},
    title = {Hot accretion flows around black holes},
    journal = {ARAA},
    volume = {52},
    pages = {529--588},
    year = {2014},
    doi = {10.1146/annurev-astro-082812-141003}
}

@article{Yuan2012,
    author = {Yuan, F. and Bu, D. and Wu, M.},
    title = {Numerical simulation of hot accretion flows: Thermal conduction effects},
    journal = {ApJ},
    volume = {761},
    pages = {130--142},
    year = {2012},
    doi = {10.1088/0004-637X/761/2/130}
}

@article{Elmegreen_1985,
	author = {Elmegreen, B. G.},
	title = {Energy dissipation in clumpy magnetic clouds},
	journal = {\apj},
	volume = {299},
	pages = {196},
	year = {1985},
	month = {dec},
	publisher = {IOP Publishing},
	doi = {10.1086/163692},
	url = {https://dx.doi.org/10.1086/163692}
}

@article{Pan2014,
    author = {{Pan}, Liubin and {Padoan}, Paolo and {Scalo}, John},
    title = "{Turbulence-induced Relative Velocity of Dust Particles. III. The Probability Distribution}",
    journal = {\apj},
    year = 2014,
    volume = 792,
    number = 1,
    pages = {69},
    doi = {10.1088/0004-637X/792/1/69},
    adsurl = {https://ui.adsabs.harvard.edu/abs/2014ApJ...792...69P},
    month = sep
}

@article{Stoll2016,
    author = {{Stoll}, Moritz H. R. and {Kley}, Wilhelm},
    title = "{Particle dynamics in discs with turbulence generated by the vertical shear instability}",
    journal = {\aap},
    year = 2016,
    volume = 594,
    pages = {A57},
    doi = {10.1051/0004-6361/201527716},
    adsurl = {https://ui.adsabs.harvard.edu/abs/2016A&A...594A..57S},
    month = oct
}

@article{Yang2024,
    author = {{Yang}, Hai and {Yuan}, Feng and {Li}, Hui and {Mizuno}, Yosuke and {Guo}, Fan and {Lu}, Rusen and {Ho}, Luis C. and {Lin}, Xi and {Zdziarski}, Andrzej A. and {Wang}, Jieshuang},
    title = "{Modeling the inner part of the jet in M87: Confronting jet morphology with theory}",
    journal = {Science Advances},
    volume = 10,
    number = 12,
    pages = {eadn3544},
    year = 2024,
    doi = {10.1126/sciadv.adn3544},
    month = mar
}

@article{King2021,
    author = {{King}, A. R. and {Pringle}, J. E.},
    title = "{Can the Blandford–Znajek Mechanism Power Steady Jets?}",
    journal = {\apjl},
    year = 2021,
    volume = 918,
    number = 1,
    pages = {L22},
    doi = {10.3847/2041-8213/ac19a1},
    eprint = {2107.12384},
    archivePrefix = {arXiv},
    primaryClass = {astro-ph.HE},
    adsurl = {https://ui.adsabs.harvard.edu/abs/2021ApJ...918L..22K},
    month = sep
}

@article{Pudritz1983,
  author = {Pudritz, R. E. and Norman, C. A.},
  title = {Centrifugally driven winds from contracting molecular disks},
  journal = {\apj},
  volume = {274},
  pages = {677},
  year = {1983},
  doi = {10.1086/161481}
}

@article{Pudritz1986,
  author = {Pudritz, Ralph E. and Norman, Colin A.},
  title = {Bipolar hydromagnetic winds from disks around protostellar objects},
  journal = {\apj},
  volume = {301},
  pages = {571-586},
  year = {1986},
  doi = {10.1086/163924}
}

@article{Xu2026,
    author    = {{Xu}, C.},
    title     = {A variable {ADAF} disk model for {X}-ray binary systems}, 
    journal   = {arXiv e-prints},
    year      = {2026},
    month     = {March},
    pages     = {arXiv:2603.10311},
    adsurl    = {https://arxiv.org/abs/2603.10311},
    archivePrefix = {arXiv},
    eprint    = {2603.10311},
    primaryClass = {astro-ph.HE}
}

@incollection{belloni2010,
    author = {Belloni, T. M.},
    title = {States and Transitions in Black Hole Binaries},
    editor = {Belloni, T.},
    booktitle = {The Jet Paradigm},
    series = {Lecture Notes in Physics},
    volume = {794},
    publisher = {Springer},
    address = {Berlin},
    year = {2010},
    pages = {},
    doi = {10.1007/978-3-540-76937-8_1}
}





\bsp	
\label{lastpage}
\end{document}